\documentclass[twocolumn]{autart}    

\usepackage{graphicx}          
\usepackage{amssymb}
\usepackage{color}
\usepackage{array}
\usepackage{subfigure}
\usepackage{latexsym}
\usepackage{amsmath}
\usepackage{color}
\usepackage{arydshln}
\usepackage{blkarray}
\usepackage{booktabs}
\usepackage{amssymb,amsmath,cite,graphicx,graphics}
\usepackage{times}
\usepackage{amssymb,amsmath,cite,graphicx,pifont}
\usepackage{stmaryrd}
\usepackage{mathrsfs}
\usepackage{amsfonts}
\usepackage{epstopdf}

\newtheorem{Lemma}{Lemma}

\newtheorem{Theorem}{Theorem}
\newtheorem{Remark}{Remark}

\begin{document}
\begin{frontmatter}

\title{Solving the Model Unavailable MARE using Q-Learning Algorithm } 


\author[a]{Fei Yan}\ead{fyan@home.swjtu.edu.cn},    
\author[a]{Jie Gao}\ead{gjie303@163.com},               
\author[a,b]{Tao Feng}\ead{sunnyfengtao@163.com},  
\author[c]{Jianxing Liu}\ead{jx.liu@hit.edu.cn}

\address[a]{School of Information Science and Technology, Southwest Jiaotong University, Chengdu, Sichuan, PR China}  
\address[b]{National Engineering Laboratory of Integrated Transportation Big Data Application Technology, Chengdu, China }  
\address[c]{Department of Control Science and Engineering, Harbin Institute of Technology, Harbin 150001, PR China} 


\begin{abstract}                          
In this paper, the discrete-time modified algebraic Riccati equation (MARE) is solved when the system model is completely unavailable. To achieve this, firstly a brand new iterative method based on the standard discrete-time algebraic Riccati equation (DARE) and its input weighting matrix is proposed to solve the MARE. For the single-input case, the iteration can be initialized by an arbitrary positive input weighting if and only if the MARE has a stabilizing solution; nevertheless a pre-given input weighting matrix of a sufficiently large magnitude is used to perform the iteration for the multi-input case when the characteristic parameter belongs to a specified subset.
Benefit from the developed specific iteration structure, the Q-learning (QL) algorithm can be  employed to subtly solve the MARE where only the system input/output data is used thus the system model is not required. Finally, a numerical simulation example is given to verify the effectiveness of the theoretical results and the algorithm.
\end{abstract}

\end{frontmatter}

\section{Introduction}
The Riccati equation constitutes an essential component within the framework of Linear Quadratic Regulator (LQR) and has been extensively investigated \cite{LQR2015,ZJF2022,Twoalgorithms}. Nevertheless, the practical considerations of imprecise detection, packet losses and stochastic disturbances cannot be disregarded, which will introduce significant uncertainties into estimation of target state. Kalman filter and optimal estimation are two typical techniques to address this problem, which are closely linked to a modified algebraic Riccati equation (MARE) that was initially derived decades ago \cite{Detection1985}.
Since then, the MARE has gained compelling attention owing to its broad applications in Kalman filtering \cite{DistributedKalmanfiltering}, intermittent observations \cite{intermittent2004}, network synchronization \cite{Stablesolution} and optimal estimator \cite{Optimalestimation}, etc. Additionally, the MARE has been further broadened to systems that are impacted by external disturbances constrained by input saturation \cite{Semi-Globaloutputconsensus}  and discrete-time mean-field systems characterized by input delays\cite{OptimalcontrolDTMS}. It has been established that the stability characteristics of such systems are intimately linked to the stabilizing solution of MARE.

Considerable amount of researches have concentrated on the solution of MARE. In the context where delectability is not assumed, the uniqueness of the almost stabilizing solution has been discussed in \cite{Almoststabilizing}. In an effort to further tackle this problem, the framework of cone-invariant operators has been employed, yielding an explicit, necessary, and sufficient condition for the existence of a mean-square stabilizing solution \cite{Existenceofamean-square}. Subsequent contributions have focused on the explicit characterization of the solution. An analytic solution for the homogeneous MARE for single-input systems \cite{Theanalyticsolutions} and a closed-form solution in terms of closed-loop poles locations \cite{Closedform} have been deduced, respectively. However, inherent requirement of solving a set of Linear Matrix Inequalities (LMIs), which requires accurate system matrices, fails to adapt these methods to model unavailable scenarios.  
In scenarios where the system model is unavailable, QL algorithm \cite{Q-learningAsma} has been employed to resolve such problems.  Subsequently, it has been implemented in multi-agent discrete-time graphical games \cite{Q-learnMulti-2014}, optimal tracking control \cite{Q-learntracking-2018} and optimal output regulation \cite{Modelfreeoutput}. In \cite{FTQ-learning}, the QL algorithm is developed to solve a single DARE, thereby facilitating the computation of the optimal feedback gain. Then the conditions for the consensus of multi-agent system is determined and the consensus is achieved when system model is completely unavailable. However, to best of our knowledge, the development of a model unavailable algorithm to solve the MARE when system model is completely unavailable remains an unsolved problem.

In this paper, we aim to develop a new iteration method to solve the MARE, and then use it to propose a model unavailable algorithm based on the QL algorithm. In \cite{OntheExistence}, the authors rewrite the MARE as a standard DARE with specific constraint and thus facilitating an iterative method. However, the initial conditions to the algorithm still hinge on the LMI problem and the iteration can only be started by trial and error, which is not suitable to the model unavailable scenario. In current study, we will address this problem by proposing a new iterative algorithm to solve the MARE. Specially, in the single-input case, we show that iteration can be started by an arbitrary positive input weighting; in the multi-input case, we reduce the selection of the input weighting matrix within the entire matrix space to a single sufficient large scalar. Based on this, the QL algorithm can be employed to solve the MARE when system matrices are completely unavailable.

  The structure of this paper is as follows. The problem is established in section 2. New iterative method to solve MARE is presented in section 3. In section 4, a model unavailable method based on QL algorithm is developed. Finally, simulation is given in section 5 and the conclusion is drawn in section 6.

  Notations: ${\lambda _i^u{\rm{(}}A{\rm{)}}}$ denotes the unstable eigenvalues of matrix $A$. ${{\lambda _{\max }}{\rm{(}}A{\rm{)}}}$ denotes the maximum eigenvalue of matrix $A$. ${\mathbb{R}^{n \times m}}$ denotes matrix space. $I_n$ denotes $n$-dimensional identity matrix. Matrix $A \succ(\succeq) B$ means matrix $A-B$ is positive (semi-) definite. $A \prec(\preceq) B$ means matrix $A-B$ is negative (semi-) definite.


\section{Problem formulation}
Consider the following discrete-time MARE
 \begin{flalign}\label{MARE2.1}
&\ X = {A^T}XA - \gamma {A^T}XB{({B^T}XB + R)^{ - 1}}{B^T}XA + Q&
 \end{flalign}
where $A$ and $B$ are system matrices, the characteristic parameter $\gamma\in (0,1)$, the input weighting matrix $R$ is positive definite  and the state weighting matrix $Q$ is semi-positive definite.
Throughout this paper, it is assumed that $(A,B)$ is stabilizable and $(A,\sqrt{Q})$ is detectable. Then, it is well known that the MARE (\ref{MARE2.1}) has a stabilizing positive definite solution when the characteristic parameter $\gamma$ is larger than
a critical value $\gamma _c$ \cite{Foundations} which is bounded by
\begin{flalign*}
  &1 - \frac{1}{{{{\max }_i}{\rm{|}}\lambda _i^u{\rm{(}}A{\rm{)}}{{\rm{|}}^2}}} \le {\gamma _c} \le 1 - \frac{1}{{\prod_i  {\rm{|}}\lambda _i^u{\rm{(}}A{\rm{)}}{{\rm{|}}^2}}}.&
\end{flalign*}

In literature \cite{OntheExistence}, the MARE can be simply solved by the LMI technique and iterative method only when the system matrices are precisely obtained. However, this paper aims to solve the MARE when the system matrices $A$ and $B$ are completely unavailable. Therefore, a new iterative method should be proposed which is especially suitable for developing a model unavailable algorithm only using the input/output date of the dynamic system.


\section{Solving MARE by a new iterative method}
In this section, we will propose a new iterative method to solve the MARE.
For the single-input case, the iterative method can be initialized by an arbitrary positive input weighting $R=r>0$ if and only if the characteristic parameter
$\gamma\in (1 - ({{\prod_i {\rm{|}}\lambda _i^u{\rm{(}}A{\rm{)}}{{\rm{|}}^2}}})^{-1},1)$, i.e., the MARE has a stabilizing solution. For the multi-input case, we restrict to
the case of $\gamma\in (\overline{\gamma},1)$ with $\overline{\gamma}$ being an estimate of ${\gamma}_c$, which facilitates us to propose the model free algorithm in the next section. It should be pointed out that the estimate $\overline{\gamma}=1-({{\prod_i {\rm{|}}\lambda _i^u{\rm{(}}A{\rm{)}}{{\rm{|}}^2}}})^{-1}$ for the single-input case.

\subsection{Single-Input Case}
For the single-input  case, the control input matrix $B=b\in \mathbb{R}^{n}$, and the input weighting matrix $R=r$ reduced to a positive scalar.
The following lemma is required for the theoretical development. 
\begin{Lemma}\label{lemma1}
Consider the following DARE
\begin{flalign}\label{DARE_s7}
    &X = {A^T}XA - {A^T}Xb{({b^T}Xb + \sigma)^{ - 1}}{b^T}XA + Q,&
 \end{flalign}
where $(A,B)$ is stabilizable and $(A,\sqrt{Q})$ is detectable. Defining that $\beta  = {b^T}Xb/\sigma$, then the scalar $\beta$ converges to
its minimum as $\sigma \to +\infty $ and the infimum is given by
  \begin{flalign*}
    &{\beta _c} = {\prod _i  {\rm{|}}\lambda _i^u{\rm{(}}A{\rm{)}}{{\rm{|}}^2}}-1.&
  \end{flalign*}
 \end{Lemma}\label{Lemma1}
 \textbf{Proof.} First, we show that $X/\sigma$ is non-increasing as $\sigma \to \infty $.
 Consider the following quadratic performance index 
  \begin{flalign*}
   &{J_\sigma}(x(0)) = x_0^T{X_\sigma}{x_0} = \sum\limits_{k = 0}^\infty  {[{x^T}(k)Qx(k) + {u^T}(k){\sigma}u(k)]}. &
  \end{flalign*}
  Suppose $0 < {\sigma_1} < {\sigma_2}$, then we will obtain
  \begin{eqnarray*}
    J_{{\sigma_1}}^*({\sigma_2}x(0))  &=& x_0^T({\sigma_2}{X_{{\sigma_1}}}){x_0} \\
      &=& \sum\limits_{k = 0}^\infty  {[{\sigma_2}{x^T}(k)Qx(k) + {\sigma_2}{u^T}(k){\sigma_1}u(k)]} \\
      &>& \sum\limits_{k = 0}^\infty  {[{\sigma_1}{x^T}(k)Qx(k) + {\sigma_1}{u^T}(k){\sigma_2}u(k)]} \\
      &\ge& x_0^T({\sigma_1}{X_{{\sigma_2}}}){x_0} = J_{{\sigma_2}}^*({\sigma_1}x(0)),
    \end{eqnarray*}
where ${X_{{\sigma_i}}}(i = 1,2)$ is the solution of the following DARE
  \begin{flalign*}
   &{X_{{\sigma_i}}} = {A^T}{X_{{\sigma_i}}}A
   - {A^T}{X_{{\sigma_i}}}b{({b^T}{X_{{\sigma_i}}}b + {\sigma_i})^{ - 1}}{b^T}{X_{{\sigma_i}}}A + Q. &
  \end{flalign*}
Since this is true for any ${x_0} \in {\mathbb{R}^n}$, the inequality ${\sigma_2}{X_{{\sigma_1}}} > {\sigma_1}{X_{{\sigma_2}}}$ holds,
that is, $({X_{{\sigma_1}}}/{\sigma_1}) > ({X_{{\sigma_2}}}/{\sigma_2})$. This indicates that ${b^T}{X_\sigma}b/\sigma$ decreases
with respect to $\sigma$. Due to the fact that ${b^T}{X_\sigma}b/\sigma > 0$, the lower bound ${\beta _c}$ exists as $\sigma \to \infty $.

Then, consider the following scalar
  \begin{flalign*}
   &\delta  = \sqrt {\frac{\sigma}{{\sigma + {b^T}Xb}}}  = \sqrt {\frac{1}{{1 + {b^T}\frac{X}{\sigma}b}}}, &
  \end{flalign*}
now we can see that the value $\delta $ reaches its maximum as ${b^T}Xb/\sigma$ reaches its minimum,
i.e., $\sigma \to \infty $. In the view of Lemma 5 in \cite{FTQ-learning}, the scalar $\delta $ has a supremum
\begin{flalign}\label{delta_c}
     &{\delta _c} =\mathop {\lim }\limits_{\sigma \to \infty } \sqrt {\frac{\sigma}{{\sigma + {b^T}Xb}}}
     = \frac{1}{{\prod_i {\rm{|}}\lambda _i^u{\rm{(}}A{\rm{)|}}}}. &
  \end{flalign}
Solving ${\beta _c}$ from equation (\ref{delta_c}), we will have
\begin{flalign}\label{beta_c}
   &{\beta _c}= \mathop {\lim }\limits_{\sigma \to \infty } \frac{{{b^T}Xb}}{\sigma}=
   {\prod_i {\rm{|}}\lambda _i^u{\rm{(}}A{\rm{)}}{{\rm{|}}^2}}-1.  &
\end{flalign}

\begin{Theorem}\label{The.1}
For the single-input MARE
 \begin{flalign}\label{mare_b}
   &X = {A^T}XA - \gamma {A^T}Xb{({b^T}Xb + r)^{ - 1}}{b^T}XA + Q,&
 \end{flalign}
there must exist a positive value ${\omega_{\gamma} }$ and a positive definite matrix ${X_\gamma }$ such that
 \begin{subequations}\label{single}
 \begin{flalign}
  &{X_\gamma } = {A^T}{X_\gamma }A -
  {A^T}{X_\gamma }b{({b^T}{X_\gamma}b + {\omega_\gamma })^{ - 1}}{b^T}{X_\gamma }A + Q,& \label{x_gamma_b}\\
 &{\omega_\gamma } = \frac{1}{\gamma }r + \frac{{1 - \gamma }}{\gamma }{b^T}{X_\gamma }b.\label{s_gamma_b}
   \end{flalign}
 \end{subequations}
 \end{Theorem}
  \textbf{Proof.}
  For any given positive initial value ${\omega_0}$, we substitute $\omega_0$ into the following
  standard algebraic Riccati equation
  \begin{flalign}\label{dare_b}
    &X = {A^T}XA - {A^T}Xb{({b^T}Xb + {\omega})^{ - 1}}{b^T}XA + Q, &
  \end{flalign}
  which yields a stabilizing solution $X_0$. If the pair $({\omega_0},{X_0})$ satisfies equation
  (\ref{x_gamma_b}) and (\ref{s_gamma_b}) already,
  then take ${\omega_\gamma } = {\omega_0}$ and ${X_\gamma } = {X_0}$, the issue is addressed. Otherwise,
  we consider the following two situations, respectively.

  1) The pair $({\omega_0},{X_0})$ satisfies equation (\ref{x_gamma_b}) such that
  \begin{flalign}\label{dayu}
     &{\omega_0} > \frac{1}{\gamma }r + \frac{{1 - \gamma }}{\gamma }{b^T}{X_0}b. &
  \end{flalign}
  Then, we construct $\omega_1$ which satisfies
  \begin{flalign*}
    &{\omega_1} = \frac{1}{\gamma }r + \frac{{1 - \gamma }}{\gamma }{b^T}{X_0}b.&
  \end{flalign*}
  Obviously, we will have ${\omega_0} \ge {\omega_1} > 0$. Similarly, substituting $\omega_1$ into DARE (\ref{dare_b})
  yields the second stabilizing solution $X_1$. It follows \cite{AREOxford}
  that ${X_0} \ge {X_1} > 0$ when $\omega_0$ is larger than $\omega_1$.
   Then, $\omega_2=r/\gamma  + [(1 - \gamma ){b^T}{X_0}b]/\gamma $
  can also be constructed and we will have two non-increasing sequences ${\omega_0} \ge {\omega_1} \ge {\omega_2} > 0$ and
  ${X_0}\succeq {X_1}\succeq {X_2} \succ 0$ for the same reason.
  Subsequently, two non-increasing sequences $\{ {\omega_0},{\omega_1},{\omega_2}, \ldots \} $
  and $\{ {X_0},{X_1},{X_2}, \ldots \}$ are easily obtained, where ${\omega_i} > 0$ and ${X_i} = X_i^T \succ 0$.
  As a result, these two sequences converge to ${\omega_\gamma }$ and ${X_\gamma }$
  that satisfies (\ref{single}).

  2) The pair $({\omega_0},{X_0})$ satisfies equation (\ref{x_gamma_b}) such that
  \begin{flalign}\label{xiaoyu}
    &{\omega_0} < \frac{1}{\gamma }r + \frac{{1 - \gamma }}{\gamma }{b^T}{X_0}b. &
  \end{flalign}
  Following the same development way in situation 1), we obtain two non-decreasing sequences $\{ {\omega_0},{\omega_1},{\omega_2}, \ldots \} $, $\{ {X_0},{X_1},{X_2}, \ldots \} $, where ${\omega_i} > 0$ and ${X_i} = X_i^T \succ 0$. Suppose sequence $\omega_i$ is bounded, these two sequences
   will definitely converge to ${\omega_\gamma }$ and ${X_\gamma }$. In the converse, i.e., the sequence $\omega_i$
    is unbounded, we will have
 \begin{eqnarray*}
  &&\mathop {\lim }\limits_{\omega_i  \to  + \infty } \frac{1}{\gamma }\frac{r}{\omega_i } + \frac{{1 - \gamma }}{\gamma }\frac{{{b^T}Xb}}{\omega_i }\\
  &=&\mathop {\lim }\limits_{\omega_i  \to  + \infty } \frac{{1 - \gamma }}{\gamma }\frac{{{\gamma _c}}}{{1 - {\gamma _c}}}\\
  &\leq& \frac{{1 - {\gamma _c}}}{{{\gamma _c}}}\frac{{{\gamma _c}}}{{1 - {\gamma _c}}}\\
  &=&1.
 \end{eqnarray*}
 due to the fact that $\gamma  \in ({\gamma _c},1]$. This indicates that with the proposed iterative method,
 there exists $\omega_i$ large enough such (\ref{dayu}) holds,
 i.e., the stabilizing solution ${X_\gamma }$ to MARE (\ref{mare_b}) is certainly to be obtained.

\begin{Remark}
In \cite{OntheExistence}, the authors have proposed a similar iterative method to solve the MARE. However,
the algorithm must started by an inequality coupling with a DARE thus the iteration can only be start
by trial and error. On the contrary, Theorem \ref{The.1} shows that the initial conditions
for the input weighting $\omega$ can be an arbitrary given positive scalar to start the iteration.
Therefore, the proposed iterative method is more ease of use. It needs to be emphasized that the crucial significance of such an improvement of the initialization of the iteration will establish the solid foundation for proposing the model free algorithm
in the following section.
\end{Remark}

\subsection{Multi-Input Case}
For the multi-input case, the input matrix $B\in \mathbb{R}^{n \times m}$ is assumed to be full column rank. The results in Lemma \ref{lemma1} are first extended to multi-input case and we will give an estimate of the critical value $\gamma_c$.

For simplicity, consider the following multi-input DARE with the input weighting matrix $\Sigma=\sigma I_m$
  \begin{flalign}\label{DARE_R}
     &X = {A^T}XA - {A^T}XB{({B^T}XB + \sigma I_m)^{ - 1}}{B^T}XA + Q,&
  \end{flalign}
 then the scalar ${\beta}={\lambda _{\max }}({B^T}XB/{\sigma}) $ which is obviously non-increasing with
respect to $\sigma$ in the view of Lemma \ref{lemma1}. Thus the infimum of $\beta$ exists which is denoted by
 ${{\beta}_c} = \mathop {\lim }\limits_{\sigma \to \infty } {\lambda _{\max }}({{B^T}XB}/{\sigma}).$ Defining that
 \begin{flalign*}
 &\bar \gamma  = \frac{{{\beta}_c}}{1 + {\beta}_c},&
 \end{flalign*}
then we have the following conclusion.
 
 \begin{Theorem}\label{The.2}
 For the DARE 
 \begin{flalign}\label{DARE_omega}
     &X = {A^T}XA - {A^T}XB{({B^T}XB + \omega I_m)^{ - 1}}{B^T}XA + Q&
  \end{flalign}
 and any given $\gamma  \in (\bar \gamma ,1)$, there exists a finite sufficient large value $\omega_t$ such that when $\omega > \omega_t$, it holds that
 \begin{flalign}\label{S_dayu}
   &\omega I_m \succ \frac{1}{\gamma }R + \frac{{1 - \gamma }}{\gamma }{B^T}XB.&
  \end{flalign}
 Furthermore, the MARE (\ref{MARE2.1}) has a stabilizing positive definite solution.
 \end{Theorem}
\textbf{Proof.} First we proof that a finite sufficient value $\omega_t$ exists for any given $\gamma \in (\bar \gamma,1)$ such that equation (\ref{S_dayu}) holds when $\omega>\omega_t$. Then an iterative method similar to single-input case can be developed when $\omega > \omega_t$.

We know that $\beta  = {\lambda _{\max }}({B^T}XB/{\omega})$ is monotonically decreasing with respect to $\omega$ and the positive definite matrices $\Omega=\omega I_m$ and $X$ satisfy (\ref{DARE_omega}). As $\omega \to \infty$, $\tilde \gamma $ converges to its infimum which is $\bar \gamma = \beta_c /(1 + \beta_c )$. Then, for any given $\gamma  \in (\bar \gamma ,1)$, there exists a finite sufficient large value $\omega_t$ such that $\gamma_t \in (\bar \gamma,\gamma)$. When taking $\omega > \omega_t$, we will obtain $1 > [(1-\gamma)/\gamma]({B^T}XB/\omega)$. Note that $\omega_t$ is sufficient large and $\omega > \omega_t$, which means $R/\omega \to 0$. Then we will have 
  \begin{flalign*}\label{1>omega}
    &1 > \frac{1}{\gamma }{\lambda _{\max }}[\frac{R}{\omega} + \frac{(1 - \gamma )}{\gamma}\frac{{{B^T}XB}}{\omega}],&
  \end{flalign*} 
which ensures the inequality (\ref{S_dayu}). Then the iterative method is developed as follows.

Construct matrix $\Omega_1$ as in \ref{The.1} and we have $\Omega\succeq{\Omega_1} \succ 0$. Substituting $\Omega_1$ into DARE (\ref{DARE_omega}), we will have the stabilizing solution $X_1$. It follows \cite{AREOxford} that $X\succeq{X_1} \succ 0$. Then $\Omega_2$, $\Omega_3$,… can be constructed in the same way and as a result, two non-increasing sequences $\{ {\Omega_0},{\Omega_1},{\Omega_2}, \ldots \} $ and $\{ {X_0},{X_1},{X_2}, \ldots \} $, where ${\Omega_i} \succ 0$ and ${X_i} = X_i^T \succ 0$, will be obtained. Obviously, these two sequences will converge to ${\Omega_\gamma }$ and $X_\gamma $ such that
 \begin{subequations}
   \begin{align*}
     {X_\gamma } &= {A^T}{X_\gamma }A -
      {A^T}{X_\gamma }B{({B^T}{X_\gamma }B + {\Omega_\gamma })^{ - 1}}{B^T}{X_\gamma }A + Q, \\
      {\Omega_\gamma } &= \frac{1}{\gamma }R + \frac{{1 - \gamma }}{\gamma }{B^T}{X_\gamma }B.
   \end{align*}
   \end{subequations}
 Then, the positive definite solution $X_\gamma$ is obtained.

  Theorem \ref{The.2} is analogous to Theorem \ref{The.1} in the previous subsection but with some constraints specifically for multi-input case.
   In a special case where matrix $\Omega$ be with a specialized structure given by $\Omega={\omega}I_m$, the MARE (\ref{MARE2.1}) admits an positive definite solution if and only if $\gamma$ is larger than the estimate value $\bar \gamma $ which can be determined via Q-Learning algorithm when input weighting matrix $R$ is characterized by a sufficient large magnitude. Given that $\omega_t$ is a finite value for any given $\gamma$ within the interval $(\bar \gamma ,1]$, one can infer that a scalar $\omega$ of sufficient large magnitude must exists, which permits the start of iteration.
  From this perspective, the QL algorithm can be adapted to solve MARE when the system model is completely unavailable.

\section{Solving MARE without system model}
In this section, the QL algorithm will be employed to derive the solution to MARE (\ref{mare_b}) when system matrices $A$ and $b$ are unavailable. In Theorem \ref{The.1}, it is established that a single standard DARE needs to be considered in each iteration. This suggests that addressing MARE equates to solve a series of DAREs. Then, a model free algorithm can be developed as follows.

 Define the $\mathbb{Q}$-function as
 \begin{eqnarray*}
\mathbb{Q}(x(k),u(k))&=&
   \left[
    \begin{array}{c}
      x(k) \\
      u(k)
\end{array} \right]^T \mathbb{H}
\left[
    \begin{array}{c}
      x(k) \\
      u(k)
\end{array} \right]\\
    &=& \left[
    \begin{array}{c}
      x(k) \\
      u(k)
\end{array} \right]^T
\left[
    \begin{array}{cc}
     \mathbb{H}_{xx}&{\mathbb{H}_{ux}^T}\\
     \mathbb{H}_{ux}&\mathbb{H}_{uu}
\end{array} \right]
\left[
    \begin{array}{c}
      x(k) \\
      u(k)
\end{array} \right],
  \end{eqnarray*}
where the kernel matrix $\mathbb{H}$ is partitioned into
\begin{flalign*}
  &{\mathbb{H}_{xx}} = Q + {A^T}XA, {\mathbb{H}_{uu}} = r + {b^T}Xb, {\mathbb{H}_{ux}} = {b^T}XA.&
\end{flalign*}
Then, we know that the optimal control is given by
\begin{flalign*}
  &u(k)=-{\mathbb{H}_{uu}^{-1} \mathbb{H}_{uu} x(k)}.&
\end{flalign*}
And $\gamma_c$ can be computed by
 \begin{flalign}
  &{\gamma_c } = 1-{\delta_c}^2 =\mathop {\lim }\limits_{r \to \infty } { \sqrt {r{({\mathbb{H}_{uu}})}^{-1}}}.&
 \end{flalign}
Therefore, the critical value $\gamma_c$ can be obtained by means of
$\delta_c$. Besides, the solution $X_\gamma$
and positive value $\omega_\gamma$ in equation (\ref{single}) can be expressed by
\begin{subequations}
\label{mubiaoH}
 \begin{flalign}
  &{X_\gamma } = {\mathbb{H}_{xx}} - \mathbb{H}_{ux}^T{{({\mathbb{H}_{uu}})}^{ - 1}}{\mathbb{H}_{ux}},&\\
  &{\Omega_\gamma } = \frac{{1 - {\gamma} }}{\gamma }{\mathbb{H}_{uu}} + r.&
 \end{flalign}
\end{subequations}
In view of this, QL algorithm emerges as a way to address a series of DAREs when system model is not accessible. The variables $\omega$ and $X$ can be computed iteratively until conditions in equations (\ref{mubiaoH}) are satisfied, i.e., the solution has been acquired.

\textbf{Algorithm 1}

\textbf{Input:} A control policy $u(k)=-Kx(k)+n(k)$, where $K$
is optimal feedback gain stabilizes $(A,b)$, $n(k)$ is an exploring noise;
a predefined small threshold $\varepsilon  > 0$; an any given initial value $\omega_0>0$.

\textbf{Step 1:} Let $Z(k) = {[{x^T}(k) {u^T}(k)]^T}$
and solve $\mathbb{H}^t$ from
\begin{eqnarray*}\label{Z{K}}
  Z(k) &=& [{x^T}(k)Qx(k)] + r{[{u^t}(k)]^T}{u^t}(k)\\
    &+&{Z^T}(k + 1){^t}Z(k + 1).
\end{eqnarray*}
\textbf{Step 2:} Obtain the iterative optimal control ${u^{t + 1}}(k)$ as
\begin{flalign*}
    &{u^{t + 1}}(k) =  - {(\mathbb{H}_{uu}^t)^{ - 1}}\mathbb{H}_{ux}^tx(k).&
\end{flalign*}
\textbf{Step 3:} Let $t \leftarrow t + 1$, repeat Step 1 and Step 2 until
 \begin{flalign*}
   &||{\mathbb{H}_{t + 1}} - {\mathbb{H}_t}|| < \varepsilon.&
\end{flalign*}
\textbf{Step 4:} Compute $\gamma_c$ by
\begin{flalign*}
   &{\gamma_c } = 1-{\delta_c}^2 =1-r{{(\mathbb{H}_{uu})}^{-1}}.&
\end{flalign*}
\textbf{Step 5:} If modified parameter $\gamma>\gamma_c$,
take $\omega_n=\omega_0$ as the initialized input weighting matrix. Otherwise, the algorithm ends.

\textbf{Step 6:} Repeat Step 1 to Step 3 and compute $X_n$ by
\begin{flalign*}
   &X_n = {\mathbb{H}_{xx}} - \mathbb{H}_{ux}^T{{({\mathbb{H}_{uu}})}^{ - 1}}{\mathbb{H}_{ux}}.&
\end{flalign*}
\textbf{Step 7:}
Obtain the iterative input weighting matrix $\omega_{n+1}$ as
\begin{flalign*}
   &\omega_{n+1} = \frac{{1 - \gamma }}{\gamma }{{{{(\mathbb{H}_{uu}}-\omega_n + r)}} + r}.&
\end{flalign*}
\textbf{Step 8:} Let $n \leftarrow n + 1$, repeat Step 6 to Step 7 until
\begin{flalign*}
   &||\omega_n-\frac{{1 - \gamma }}{\gamma }{\mathbb{H}_{uu}} + r|| < \varepsilon.&
\end{flalign*}
\textbf{Step 9:} Obtain the solution $x_\gamma=x_n$.

\textbf{End.}

\begin{Remark}
As  $t \to  + \infty $, it is established in \cite{Q-learningAsma} that the QL algorithm converges. And the iterative kernel matrix $\mathbb{H}$ will converge to the true value. Assuming the input weighting matrix $r$ be sufficiently large at first, the critical value $\gamma_c$ in single-input case can be approximated with an arbitrary degree of precision.
Therefore, the existence of stabilizing solution for MARE (\ref{mare_b}) can be verified when system matrices $A$ and $b$ are entirely unknown.
\end{Remark}

\begin{Remark}
It is evident that an error is introduced at each iteration step. When implementing QL algorithm, the error is accumulated iteratively, potentially leading to the algorithm's failure to converge particularly in high-dimensional problems. In light of Theorem \ref{The.1}, it is observed that if the pairs $({\omega_l},{X_l}{\rm{)}}$ and $({\omega_r},{X_r}{\rm{)}}$ satisfy inequality (\ref{xiaoyu}) and (\ref{dayu}) respectively, the two sequences ${X_l} \prec {\rm{ }}{X_\gamma } \prec {X_r}$, ${\omega_l} \le {\rm{ }}{{\rm{\omega}}_\gamma } \le {\omega_r}$ holds.
Hence, it suffices to consider the appropriately selected parameter $\omega$. Only a single DARE (\ref{x_gamma_b}) needs to be solved until the condition (\ref{s_gamma_b}) is met. The dichotomy method based on QL algorithm can also be developed which ensures a higher accuracy.
\end{Remark}


\section{Simulation}
The system matrices are given by
\begin{flalign*}
&A=\left[
    \begin{array}{cc}
      1 & 0.5 \\
      0 & 1 \\
    \end{array}
  \right],\ \ b=\left[
                  \begin{array}{c}
                    0 \\
                    1 \\
                  \end{array}
                \right].&              \label{ABform}
\end{flalign*}
Obviously, matrix $A$ is neutrally unstable.
Set $Q = {I_2}$, $r = 10$, $\gamma=0.8$.
Take ${\omega_0}  = 10$ as the initial value, then we compute ${\omega_\gamma}$
and ${X_\gamma }$ by the iterative method in Theorem \ref{The.1} and
after 7 iterations, we obtain $X_\gamma$ and $\omega_\gamma$ as follows
\begin{eqnarray*}
 {X_\gamma }&=&\left[
    \begin{array}{cc}
      5.5114 & 5.2161 \\
      5.2161 & 11.7659 \\
    \end{array}
    \right],\\
  {\Omega_{\gamma}} &=& 15.4415
  \end{eqnarray*}
within the error of  ${10^{-7}}$.

When the system matrices are unavailable, we run Algorithm 1 and obtain the solution to MARE (\ref{mare_b}). Fig.1 shows the convergence process of ${\gamma_c}^t$, where ${\gamma_c}^{120}=0.1653$. As we can see in Fig.2, the kernel matrix $\mathbb{H}$ converges to the true value with the error less than $10^{-10}$ at each iteration. For each iteration, we compute ${X_n}$ and the norm of ${\omega_n } - [(1 - \gamma )/\gamma ]{\mathbb{H}_{uu}} - r$ which is presented in Fig.3. As shown in Fig.3, the error is less than $10^{-12}$ after 60 iterations. Let $\Delta(X)={A^T}XA -\gamma {A^T}Xb{({b^T}Xb + r)^{ - 1}}{b^T}XA + Q$, the norm of $X_n-\Delta_n(X_n)$ is presented for the $n$th iteration in Fig.4 which can be used to determine the solution. When the difference is $0$, $X_n=\Delta_n(X_n)$ and MARE (\ref{mare_b}) holds, which means the solution is obtained by $X_\gamma=X_n$. After 60 iterations, the solution ${X_\gamma}$ and the positive value ${\omega_\gamma}$ are obtained as
\begin{eqnarray*}
  {X_\gamma } &=&\left[
    \begin{array}{cc}
      5.5114 & 5.2161 \\
      5.2161 & 11.7659 \\
    \end{array}
    \right],\\
    {\omega_\gamma } &=& 15.4415,
\end{eqnarray*}
with the error $\varepsilon  = {10^{ - 12}}$ achieved, which means the
 MARE is solved by Algorithm 1.


 \begin{figure}\label{Fig.gamma}
  \begin{center}
  \includegraphics[height=3.8cm]{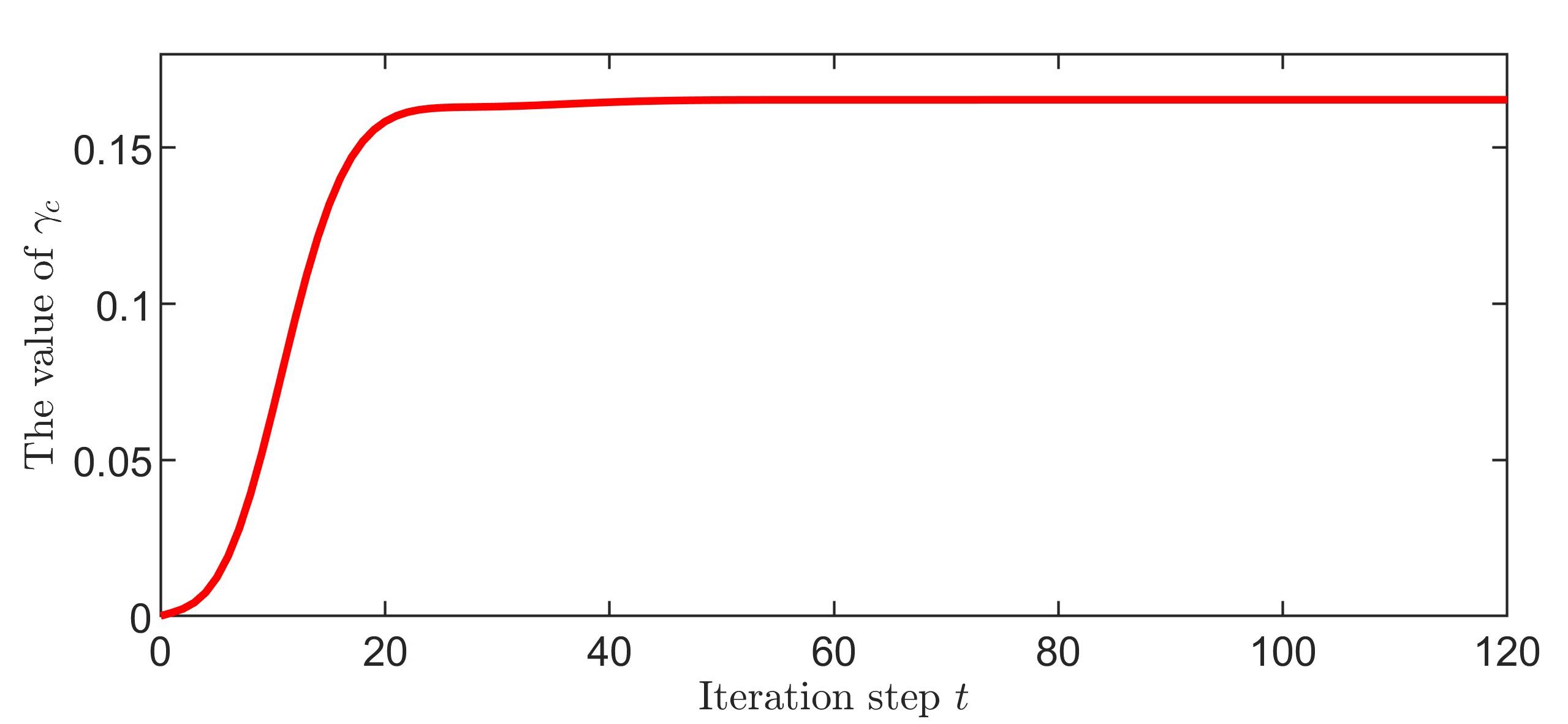}
  \caption{Convergence process of $\gamma_c$}
  \label{Q-learning-gamma}
  \end{center}
\end{figure}

 \begin{figure}\label{Fig.1}
  \begin{center}
  \includegraphics[height=3.8cm]{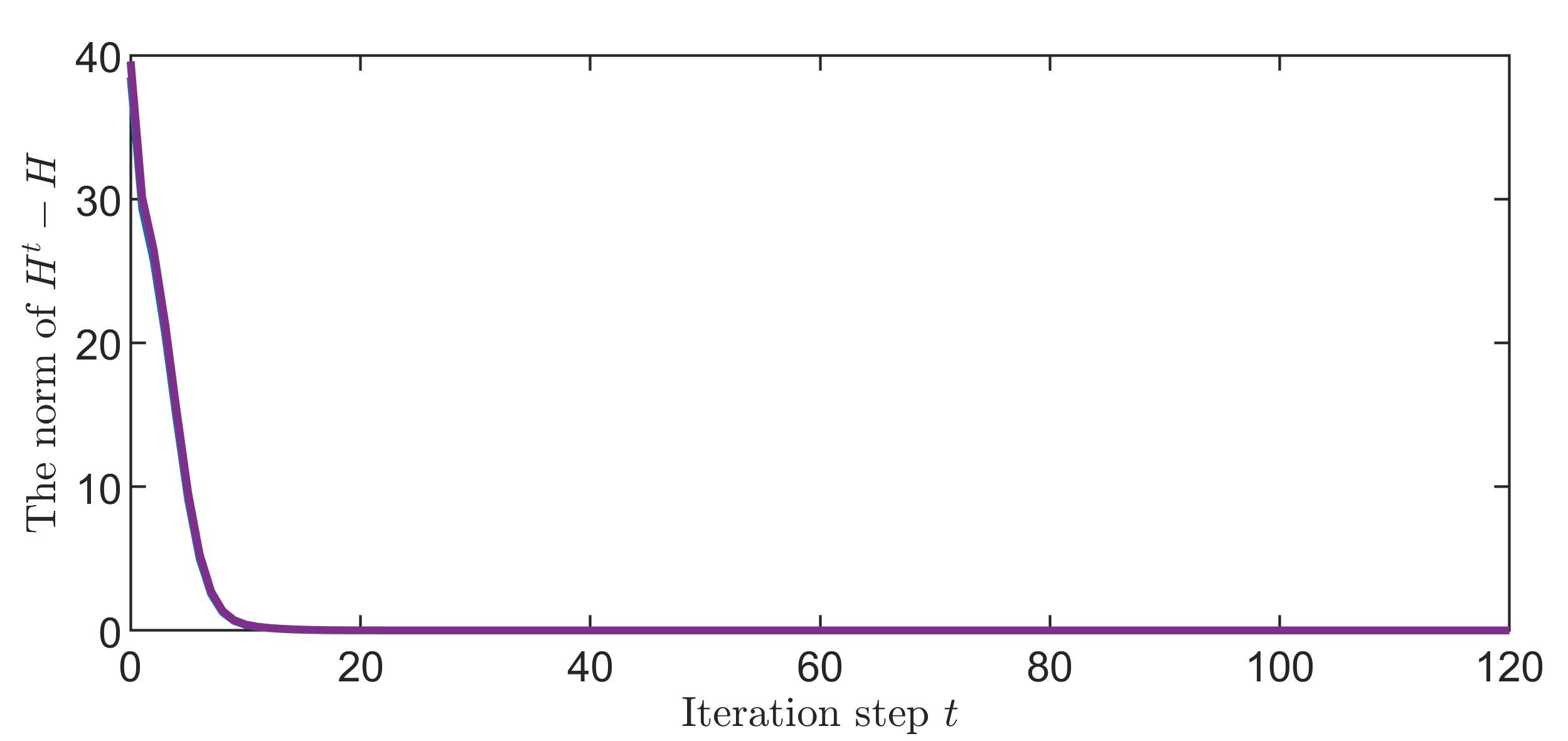}
  \caption{Error between matrix $\mathbb{H}^{t}$ and the true value of $\mathbb{H}$}
  \label{Q-learning-H}
  \end{center}
\end{figure}

\begin{figure}\label{Fig.2}
  \begin{center}
  \includegraphics[height=3.8cm]{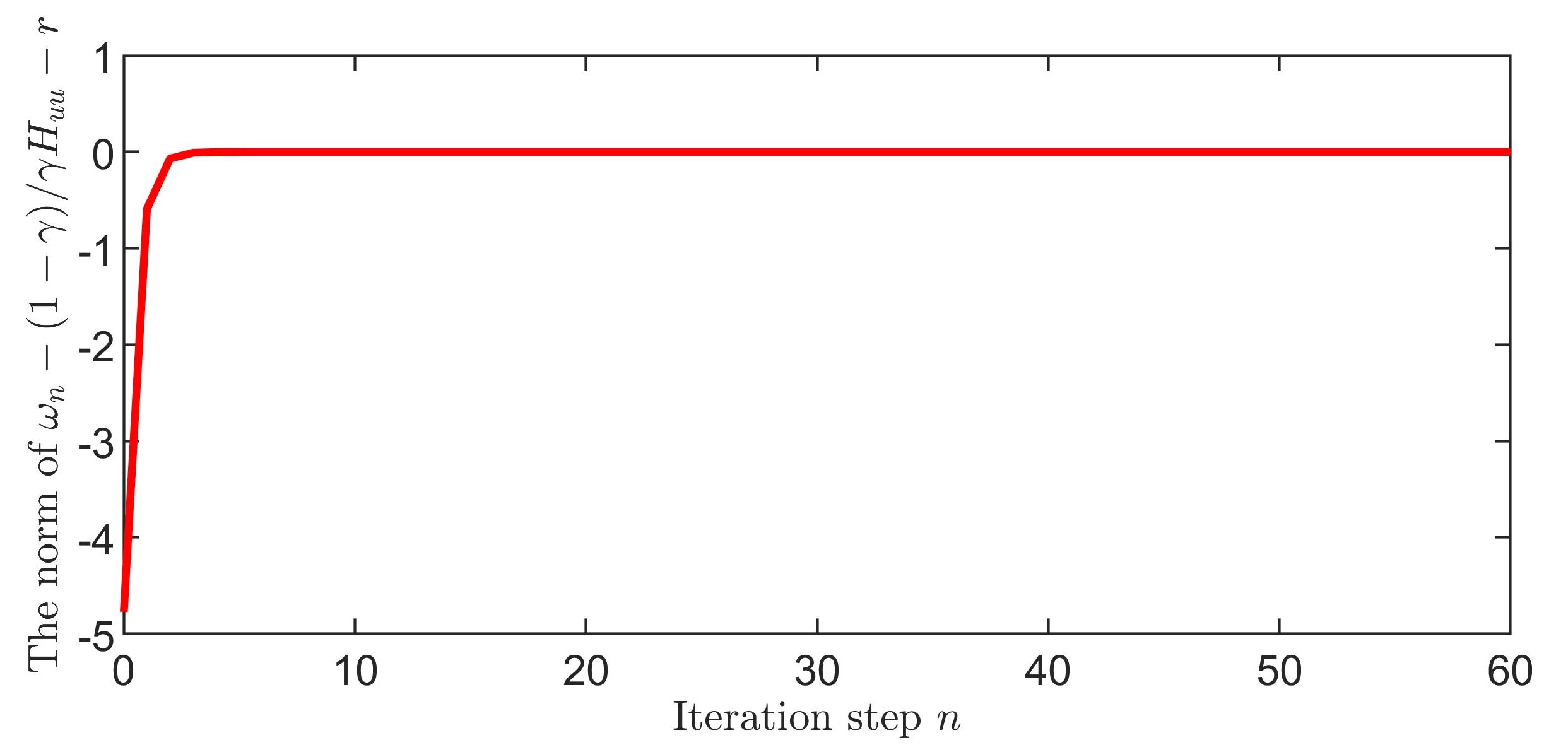}
  \caption{The norm of ${\omega_n} - \frac{{1 - \gamma }}{\gamma }{\mathbb{H}_{uu}} - r$}
  \label{Q-learning-CMP}
  \end{center}
\end{figure}

 \begin{figure}\label{Fig.3}
  \begin{center}
  \includegraphics[height=3.8cm]{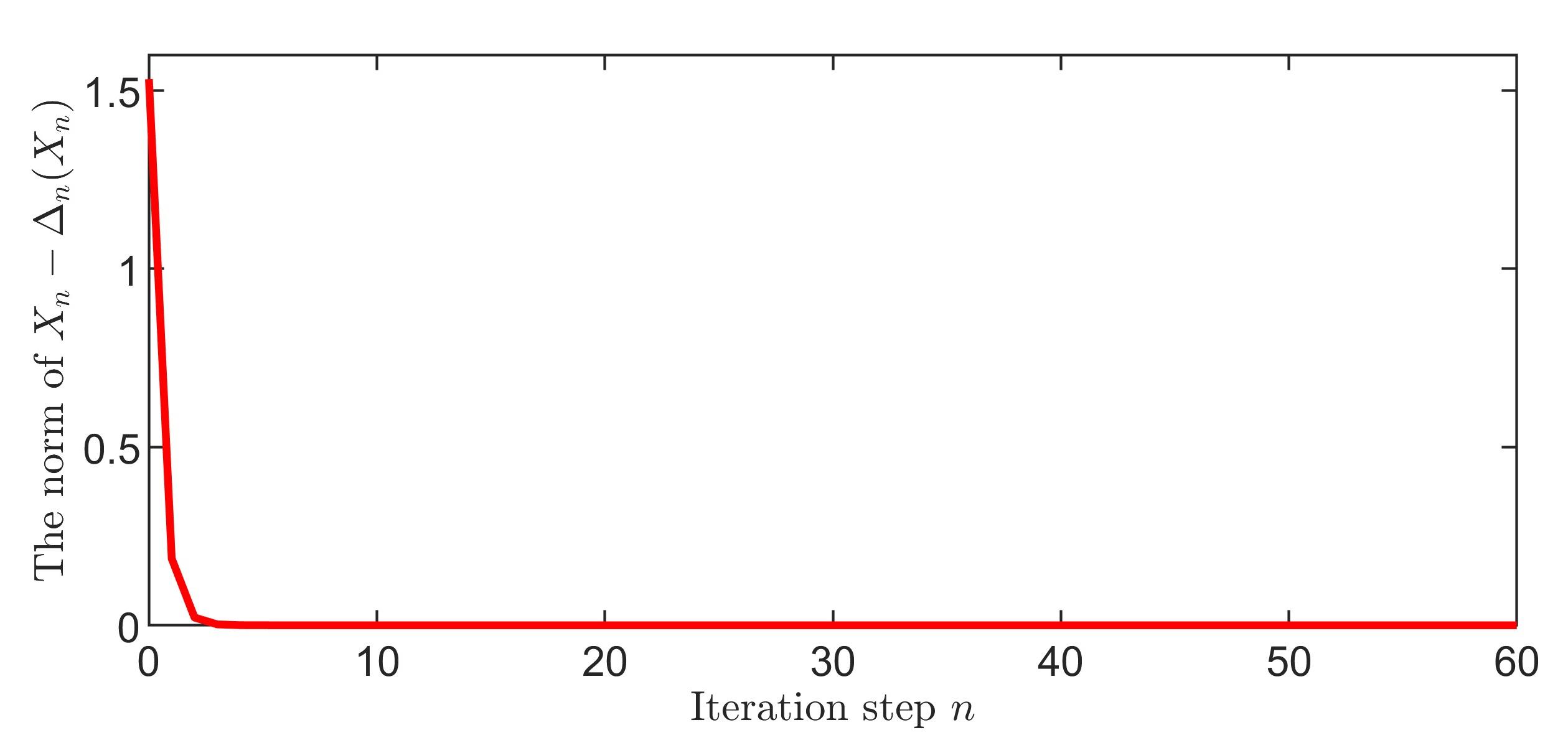}
  \caption{The norm of $X_n-\Delta_n (X_n)$ }
  \label{Q-learning-MARE}
  \end{center}
\end{figure}


\section{Conclusion}
In current study, the modified algebraic Ricatti equation (MARE) is solved when system model is absolutely unavailable. A novel iterative approach is developed and explicated for both single-input and multiple-input case. Furthermore, an estimate of the critical value $\gamma_c$ is deduced by our proposed method, which is attainable when system model is unavailable via QL algorithm.
For single-input case, the positive definite solution of MARE can be obtained from arbitrarily specified input weighting value. Moreover, a dichotomy method is introduced to guarantee the accuracy of the solution.
For multi-input case, the results are presented for a particular instance, $\Omega=\omega I_m$. Upon the pre-given positive definite matrix $\Omega$ with sufficiently large parameter $\omega$, the problem is tackled when the characteristic parameter $\gamma$ is larger than the pre-determined estimate bound $\bar \gamma $.
Based on this, a model-free algorithm is developed and by which the MARE is solved.

\bibliographystyle{unsrt}       
\bibliography{QL_mare} 
\end{document}